\documentclass{cidr-2020}
\usepackage{times}
\usepackage{microtype}
\usepackage{etex}
\usepackage{booktabs} %
\usepackage[usenames, dvipsnames]{color}
\usepackage{xcolor}
\usepackage{balance}
\usepackage{url}
\usepackage{graphicx}
\usepackage{array}
\usepackage{multirow}
\usepackage{mathtools,amssymb}
\usepackage[ruled,vlined,linesnumbered]{algorithm2e}
\usepackage{verbatim}
\usepackage{comment}
\usepackage[normalem]{ulem}
\usepackage{enumitem}
\usepackage{booktabs}
\usepackage[colorlinks=true,citecolor=magenta,linkcolor=magenta,urlcolor=magenta]{hyperref}
\usepackage{soul}
\usepackage{multirow}
\usepackage{multicol}
\usepackage{flushend}
\usepackage{epsfig}
\usepackage{endnotes}
\usepackage{subfig}
\usepackage{enumitem}
\usepackage{epstopdf}

\def\Snospace~{\S{}}
\def\Fnospace~{\mbox{Fig.\hspace{0.25em}}}
\def\Tnospace~{\mbox{Tab.\hspace{0.25em}}}
\def\Enospace~{\mbox{Equation\hspace{0.25em}}}

\newcommand{\eg}{e.g.,\ }

\newcommand{\vspacebfigure}{\vspace{-2mm}}
\newcommand{\vspaceafigure}{\vspace{-2mm}}

\definecolor{lightgray}{rgb}{.85,.85,.85}  %
\definecolor{orange}{RGB}{255,127,0}
\sethlcolor{lightgray} %

\newcommand{\sstitle}[1]{\noindent\textbf{#1}}

\newcommand{\eat}[1]{}

\def\compactify{\itemsep=0pt \topsep=0pt \partopsep=0pt \parsep=0pt}
\let\latexusecounter=\usecounter

\newcommand{\sysname}{Flock\xspace}

\begin{document}

\title{Cloudy with High Chance of DBMS:\\ A 10-year Prediction for Enterprise-Grade ML}

\author{
  \alignauthor
	{\fontsize{10}{11} \selectfont Ashvin	Agrawal, Rony	Chatterjee,
Carlo	Curino, Avrilia	 Floratou,
Neha	Gowdal, Matteo	Interlandi, 
Alekh	Jindal,~~~~~ \\ Konstantinos	 Karanasos,
Subru	Krishnan, Brian	Kroth, 
Jyoti	Leeka, Kwanghyun	Park, 
Hiren	Patel,  Olga	 Poppe,~~~~~\\
Fotis	Psallidas, 
Raghu	Ramakrishnan,
Abhishek	 Roy, Karla	Saur,
Rathijit 	Sen,  Markus	Weimer,
Travis	Wright, Yiwen	Zhu~~~~~      \\[2mm]
      \email{{\em name.surname@microsoft.com}}\\[1mm]
      \affaddr{Microsoft}
      }
}

\maketitle

\begin{abstract}
Machine learning (ML) has proven itself in high-value web
applications such as search ranking and is emerging as a powerful tool in a much broader range of  enterprise scenarios including voice recognition and conversational understanding for customer support, autotuning for videoconferencing, intelligent feedback loops in large-scale sysops, manufacturing and autonomous vehicle management, complex financial predictions, just to name a few.

Meanwhile, as the value of data is increasingly recognized and monetized, concerns about securing valuable data and risks to individual privacy have been growing. Consequently, rigorous data management  has emerged as a key requirement in enterprise settings.  

How will these trends (ML growing popularity, and stricter data governance) intersect? What are the unmet requirements for applying ML in enterprise settings? What are the technical challenges for the DB community to solve?  In this paper, we present our vision of how ML and database systems are likely to come together, and early steps we take towards making this vision a reality.
\end{abstract}

\section{Introduction}
\label{sec:intro}

Machine learning (ML) has proven itself in high-value consumer 
applications such as search ranking, recommender systems and spam 
detection~\cite{yahoo-spam, yahoo-coke}. These applications are built and 
operated by large teams of experts, and run on massive dedicated 
infrastructures.\footnote{ML.NET~\cite{ml.net} alone took dozens of 
engineers over a decade.} The (exorbitant) human and hardware costs are 
well justified by multi-billion dollar paydays. This approach is 
entirely impractical when it comes to the (ongoing) mainstream adoption 
of ML.

\interfootnotelinepenalty=10000

Enterprises in every industry are developing strategies for digitally 
transforming their business at every level. The core idea is to 
continuously monitor all aspects of the business, actively interpret the 
observations using advanced data analysis---including ML---and integrate 
the learnings into appropriate actions that improve business outcomes. We 
predict that in the next 10~years, hundreds of thousands of small teams 
will build millions\footnote{From our analysis on \textgreater4M Python notebooks from public GitHub repositories~\cite{dsonds}, we conservatively estimate that 10\% of the world's developers will use ML in
the next 10 years---totalling 20M engineering years.} of ML-infused applications---most just moderately remunerative, 
but with huge collective value.   

When it comes to leveraging ML in {\em enterprise} applications, 
especially in regulated environments, the level of scrutiny for data 
handling, model fairness, user privacy, and debuggability will be 
substantially higher than in the first wave of ML applications. Consider 
the healthcare domain: ML models may be trained on sensitive medical data,
and make predictions that determine patient treatments---copying CSV 
files on a laptop and maximizing {\em average} model accuracy just 
doesn't cut it! We refer to this new class of applications as {\em Enterprise Grade Machine Learning (EGML).}

In this paper, we speculate on how ML and database systems will evolve to 
support EGML over the next several years. Database management systems 
(DBMSs) are the repositories for high-value data that demands security, 
fine-grained access control, auditing, high-availability, etc. Over the last 30~years, whenever 
a new data-related technology has gained sufficient adoption, inevitably 
DBMS vendors have sought to absorb the technology into their mainstream 
products. Examples include Object-Oriented~\cite{oodbms}, XML~\cite
{xmldbms}, and Big Data~\cite{SQLServer2019} technologies. Indeed, ML is 
no exception if we consider SQL Server Analysis Services~\cite{ssas}, SQL 
Server R and Python integration~\cite{ssml}, and Big Query support for 
ML~\cite{bigquery-ml}. {\em Is the future then that ML will be 
assimilated by the  DBMS?}

 \begin{figure*}
	\centering\includegraphics[width=1.9\columnwidth]{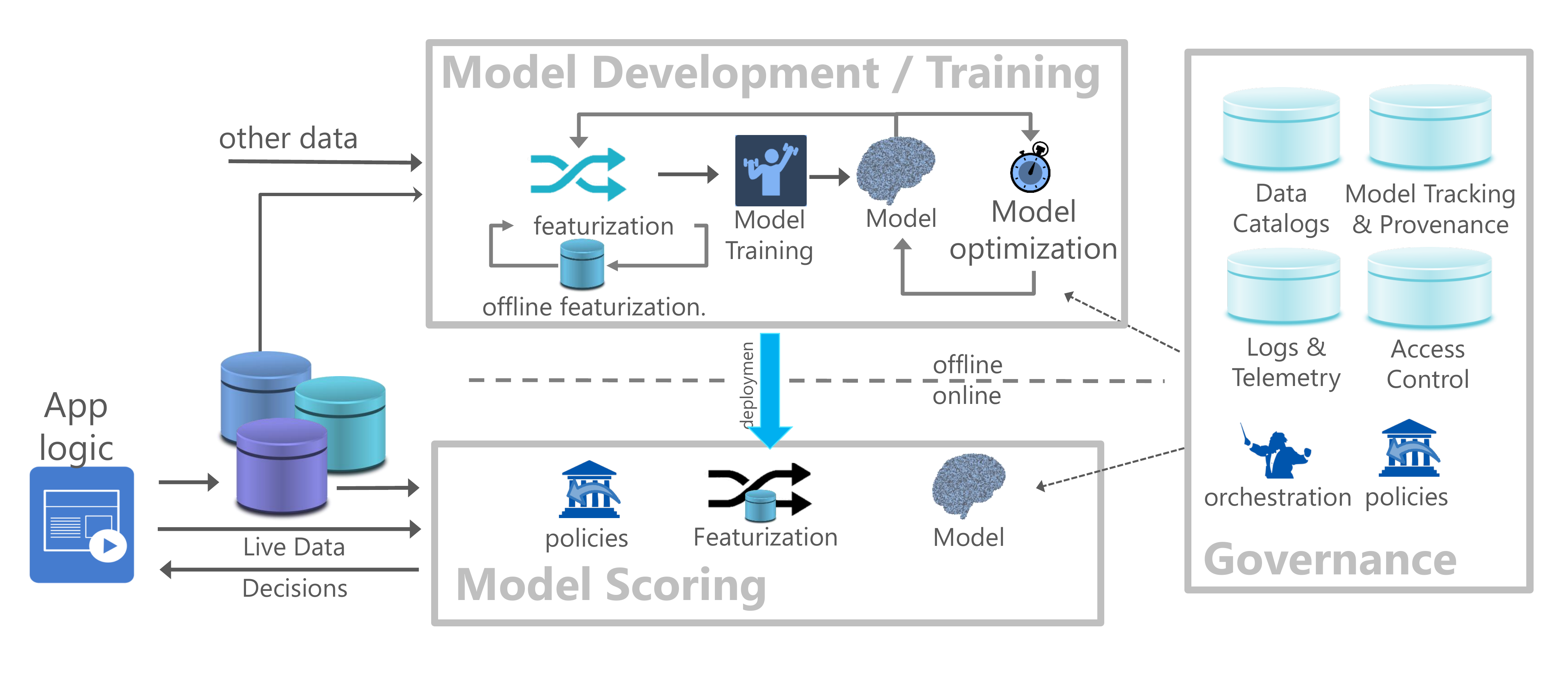}
        \vspacebfigure
 	\caption{\sysname reference architecture for a canonical data science lifecycle. }
        \label{fig:flock}
        \vspaceafigure
 \end{figure*}

We believe that this is too simplistic, and understanding the path forward
requires a more careful look at the various aspects of EGML, which we divide
into three main categories: {\em model development/training}, {\em model
scoring} and {\em model management/governance}.

\indent \sstitle{Train in the Cloud.} First, we are witnessing an ongoing
revolution in frameworks for training an increasingly broad range of ML model
classes. Their very foundations are still undergoing rapid
development~\cite{chen2018neural}. Often, these developments happen in
conjunction with innovations in hardware. The rapidly expanding community of
data scientists who train models are developing sophisticated environments for
managing and supporting the iterative process of data exploration, feature
engineering, model training, model selection, model deployment, etc.---prominent examples include~\cite{mlflow, azureml}. These large, complex, evolving infrastructures are a
good fit with managed cloud service infrastructure. Moreover, model training
requires centralized data, is characterized by spiky resource usage, and
benefits from access to the latest hardware. This leads us to believe that model
training and development will happen in either private or public clouds.

\indent \sstitle{Score in the DBMS.} Second, while the models may be centrally
trained the resulting inference pipelines will be deployed everywhere: in the
cloud, on-prem, and on edge devices to make inferences (``scoring'') where the
data is. This raises the question of whether doing inference on data stored in a
DBMS can be done as an extension of the query runtime, without the need to
exfiltrate the data. We strongly believe this can and must be supported. It
appears likely that the most widely studied or promising families of models can
be uniformly represented \cite{mlflow,onnx}. Given a particular model we can
express how to score it (i.e., perform inference) on an input using an appropriate algebra, and compile these algebraic structures into highly optimized code for different
execution environments and hardware~\cite{tvm}.  Taken together, these
observations suggest that we need to consider how to incorporate ML scoring as a
foundational extension of relational algebra, and an integral part of SQL query
optimizers and runtimes---we present a concrete proposal in
\S~\ref{sec:inference}.

\indent \sstitle{Governance everywhere.} Third, we believe that all data, 
including deployed models---models are, in fact, best thought of as 
derived data---and  the inferences made using them, will need to be 
robustly governed. The deployment of ML models and their use in 
decision making via inference 
leads to many significant challenges in governance. For example, 
regulations such as GDPR~\cite{gdpr} and concerns such as model bias and 
explainability motivate tracking provenance all the way from data used 
for training through to decisions based on scoring of trained 
models. In turn, this requires efficient support for versioning data. 
While the ML community is focused on improvements in algorithms and 
training infrastructures, we see massive need for the DB community to 
step up in the areas of secure data access, version management, and 
provenance tracking and governance---we discuss initial work in this area 
in \S~\ref{sec:datamanagement}. 

In summary, the future is likely {\em cloudy with a high chance of DBMS, and governance throughout.} We describe how our vision is shaped by customer conversations, data and market analysis, and our direct experience as well as present several open problems. We conclude by highlighting promising initial results from a few of the solutions that we are working on.

\section{The Flock vision}

In this section, we present our vision for {\em \sysname}, a reference architecture to support the canonical data science lifecycle for EGML applications. {\em Flock} is our vehicle to explore assumptions (\S~\ref{sec:vantagepoint}), discover open problems and validate initial solutions (\S~\ref{sec:problems}).
We start from a key observation: {\em Machine Learning models are software artifacts derived from data.} The resulting dual nature of software artifacts and derived data provides us with a useful lens to understand the role of the DB community in the EGML revolution.

The lifecycle shown in Figure~\ref{fig:flock} begins with a (typically) offline phase, where a data scientist (and more and more frequently any software engineer) gathers data from multiple data sources, transforms them via several featurization steps and models reality using learning algorithms. Today, this phase is very manual and sadly closer to a black art than an engineering discipline. Looking at ML as software, we expect the ML and Software Engineering communities to provide us with automation~\cite{automl_book}, tooling, and engineering best practices---ML will become an integral part of the DevOps lifecycle.  Looking at ML models as derived data, the DB community must address data discovery, access control and data sharing, curation, validation, versioning and provenance~(\S~\ref{sec:datamanagement}). Moreover, today's prevalent abstraction for data science is imperative Python code orchestrating data-intensive processing steps, each performed within a native library. This suggests that end-to-end ML pipelines can be approached as inherently optimizable dataflows (\S~\ref{sec:inference}).

The second stage in the lifecycle is entered when a model is selected and is ready to be deployed. Using the models-as-software lens, deployment consists of packaging the entire inference pipeline (model + all data preprocessing steps) in a way that preserves the exact behavior crafted by the data scientist in the training environment. Next, deployment envolves finding a suitable hosting infrastructure for scoring the model. Today's best practice is to package models in costly containers and hope that enough of the environment is preserved to ensure correctness\footnote{This is optimistic (e.g., is floating point precision guaranteed when running a container across Linux/Windows, x64/ARM?)}. Recall that in EGML settings individual decisions could be very consequential (e.g., loan acceptance, or choice of medical treatment), so ``average model accuracy'' is not a sufficient validation metric.   Switching to our models-as-data lens, we observe that they must be subject to GDPR-style scrutiny, and their storage and querying/scoring must be secured and auditably tracked---e.g., \cite{carlini2018secret} has shown how to extract credit card numbers memoized in a trained neural network language model. Also, privacy and fairness implications must be handled carefully. Moreover, as the underlying data evolves, models need to be updated. To retain consistency for complex applications multiple models might have to be updated transactionally. DBMSs have long provided these type of enterprise features for operational data, and we propose to extend them to support model scoring. While this was our primary motivation, our early experiments suggest that in-DB model scoring actually allows us to deliver $5\times$ to $24\times$ speedups over standalone state of the art solutions!

Model predictions usually come in the form of single numbers or vectors of numbers (\eg the probability of each class in a classification problem). To act on a prediction, it must be transformed to domain terms (\eg the name of the winning class). But actions are typically more nuanced and involve policies that encode business constraints and might actually override a model's prediction under certain circumstances. Systematizing this policy space is important, as we have discussed in~\cite{floratou2017dhalion}.

Beside basic orchestration (of what happens when) in project lifecycle,  management and governance for data and models throughout is vital.  Access to a deployed model must be controlled, similar to how access to data or a view is controlled in a DBMS. Provenance here plays a key role and has two distinct applications---looking at models as software artifacts, we must be able to verify them or debug them, even as they evolve due to re-training. From a model-as-data viewpoint, we must be able to determine how a model was derived and from which snapshot of (training) data, in order to interpret the predictions and answer questions such as whether and why certain decisions were biased. 
This leads to the need for pervasive and automated tracking of provenance from training through deployment to scoring (\S~\ref{sec:datamanagement}). 

Given this context, we argue that: 1) the {\em ML development/training will happen in the Cloud}; 2) {\em Models must be stored and scored in managed environments such as a DBMS}; and 3) {\em Provenance needs to be collected across all phases}.

\section{The vantage point}
\label{sec:vantagepoint}

\noindent Our perspective on {\em what Enterprise-grade ML (EGML) will look like in 10 years} is shaped by multiple inputs. 

\vspace{2mm}
{\noindent \em \large First-hand experience.} Collectively, the authors of this paper have extensive experience in using ML technologies in production settings, e.g., content recommenders~\cite{yahoo-coke}, spam filters~\cite{yahoo-spam}, big data learning optimizers~\cite{costLearner,cardLearner,raqo-full,raqo}, ML-based performance debuggers, Azure cloud optimizations based on customer load predictions, self-tuning streaming systems, and auto-tuning infrastructures for SQL~Server internals. Many of us have also been working on systems for ML technologies, including big data infrastructure \cite{hydra,hadoop,netco}, ML toolkits~\cite{ml.net,nimbus-ml,onnx-runtime}, and the systems that orchestrate it all in the cloud~\cite{azureml}. 

This experience has led to one key insight: {\bf ``An ML model is software derived from data''}. This means that ML presents characteristics that are typical of software (e.g., it requires rich and new CI/CD pipelines), and of data (e.g., the need to track lineage)---hence, the database community is well positioned to play a key role in EGML, but much work is needed. We discuss some of the problems we tackle in \S~\ref{sec:problems}.

A second---and painfully clear---observation is that {\bf the actual model development represents less than 20\% of most data science project lifecycle}~\cite{kagglesurvey}. The majority of the time is spent on collecting and cleaning the data, with the reminder on operationalizing the best model.

\vspace{2mm}
{\noindent \em \large Conversations with enterprises.}
We have engaged with many large, sophisticated enterprises, including: (i) a financial institution seeking to streamline its loan approval process, (ii) a marketing firm identifying which customers to target for promotions, (iii) a sports company predicting athletes' performance, (iv) a health insurance agency aiming to predict patient recidivism, and (v) a large automotive company modeling recalls, customer satisfaction, and marketing.

A key learning from these conversations is that compared to ``unicorn'' ML applications like web search, these enterprise applications are characterized by smaller teams with domain expertise rather than deep algorithmic or systems expertise. On the other hand, their platform requirements are much more stringent around auditing, security, privacy, fairness, and bias.\footnote{This is not intended to suggest that unicorn applications do not share these requirements; rather, enterprise teams want off-the-shelf platforms that have a much higher level of support built-in, whereas unicorn teams have typically built everything from scratch.} This is particularly true for regulated industries. Existing ML technologies are not ready to support these applications in a safe, cost-effective manner. We also learned that is increasingly common for a data science project to produce a large number of localized/specialized models, rather than a single uber model---e.g., a scam-detection model per ATM vs a single global model. This creates novel challenges on model management, deployment, tracking, (transactional) rolling forward/backward among model versions etc. In short:  {\bf Typical applications of ML are built by smaller, less experienced teams, yet they have more stringent demands}.

\begin{figure}
	\centering\includegraphics[width=0.9\columnwidth]{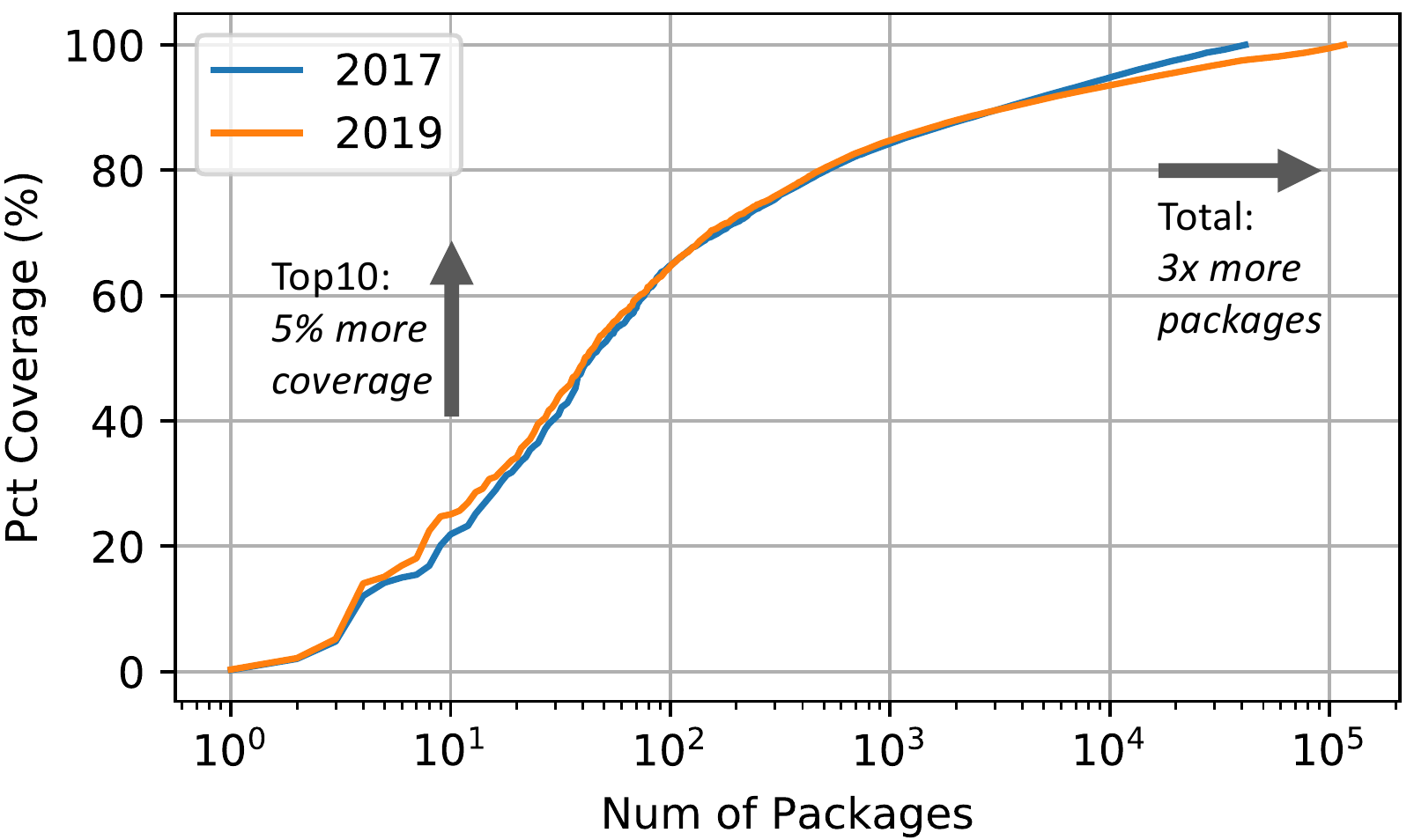}
        \vspacebfigure
	\caption{Notebook coverage (\%) for top-K packages.}
	\label{fig:coverage}
        \vspaceafigure
\end{figure}

\vspace{2mm}
{\noindent \em \large GitHub analysis.}
To get a feel for trends in the broader data science community, we {\em downloaded and analyzed $>4$	 million public Python notebooks from GitHub}, plus hundreds of thousands of data science pipelines from within Microsoft~\cite{dsonds}. Over 70\%  of the notebooks focused on ML. Moreover, we analyzed hundreds of versions for popular Python packages. The details of this analysis are beyond the scope of this paper, but we make a few key observations.  Figure~\ref{fig:coverage} shows the fraction of notebooks that would be completely supported, if we only covered the K most popular packages (for varying values of K). The shift between 2017 and 2019 suggests that the field is still expanding quickly (many more packages) but also that we are seeing an initial convergence (a few packages are becoming dominant). For example, \texttt{numpy}, \texttt{panda}s and \texttt{sklearn} are solidifying their position. We also observe very limited adoption of solutions for testing/CI-CD/model tracking (MLFlow~\cite{mlflow} is still not very popular despite its relevance to EGML).

Overall this suggests that {\bf systems aiming to support EGML must provide broad coverage, but can focus on optimizing a core set of ML packages.} 

\begin{figure}
	\centering\includegraphics[width=\columnwidth]{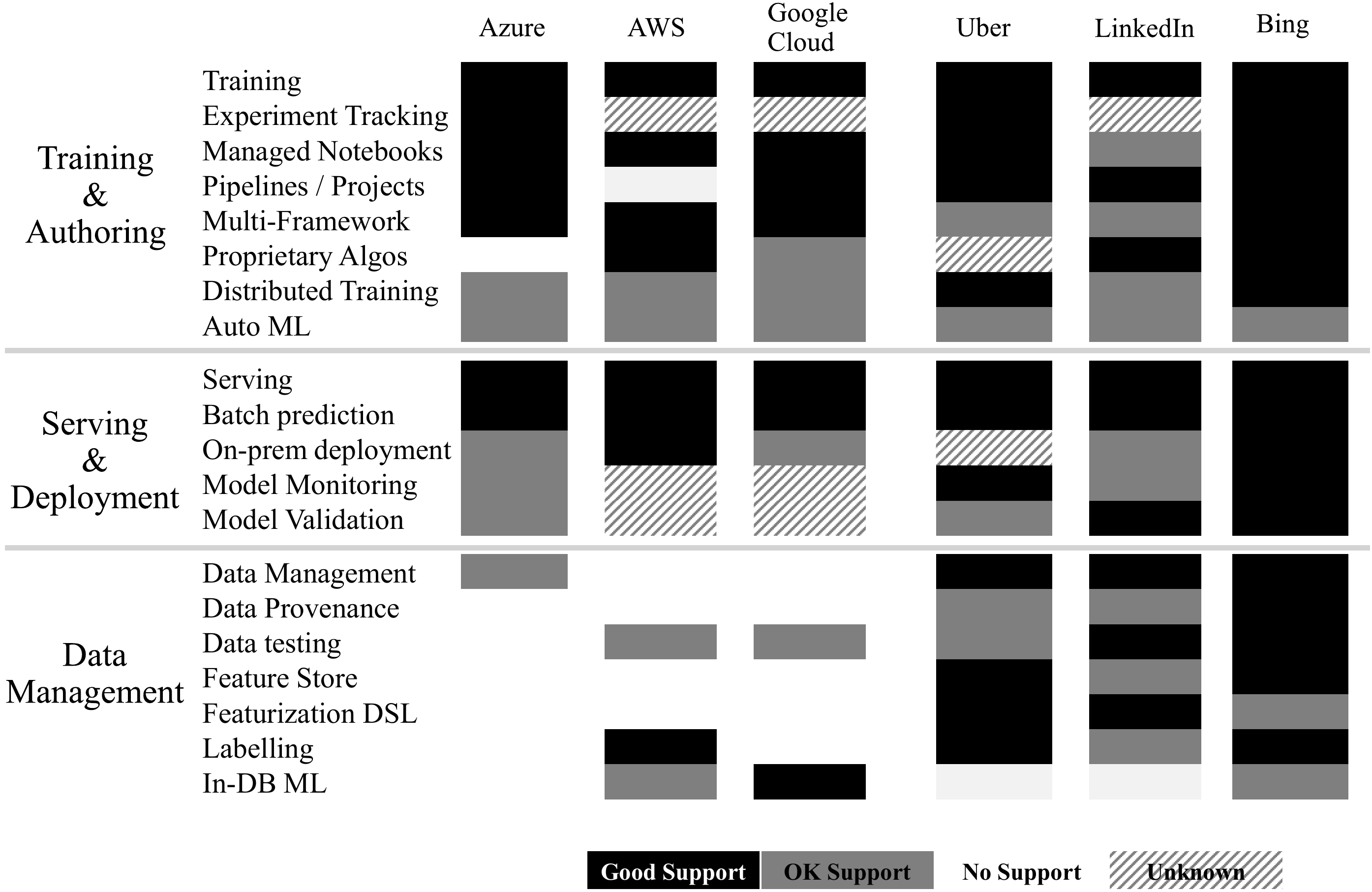}
	\vspacebfigure
	\caption{ML Systems in the public cloud and major companies.}
	\label{fig:relatedwork}
	\vspaceafigure
\end{figure}

\vspace{2mm}
{\noindent \em \large Competitive landscape.} Many of the companies that built the first ``unicorn'' ML applications also developed systems to support the data science lifecycle.  Some of those tech stacks were later released as open source while others lead to services on the public cloud. In Figure~\ref{fig:relatedwork}, we compare some of the most mature systems in this area (with accessible information), based on the level to which they support different features\footnote{This is ostensibly a subjective judgement based on a few weeks of analysis of marketing material, code skimming, and light experimentation.}. Note that the area is dynamic and that the table reflects our understanding of these systems at the time of writing. We consider Bing, Uber~\cite{uber} and LinkedIn's ProML~\cite{proml} as examples of proprietary infrastructures powering ``unicorn'' applications.  Also, all major public cloud providers have services to support enterprise machine learning~\cite{azureml, googlecloudml, sagemaker}.

Analyzing those systems, we identified key feature areas: Training, Deployment and Data Management. A detailed discussion is beyond the scope of this submission.
But we observe two major trends: 1) {\bf mature proprietary solutions have stronger support for data management}---this is consistent with our own direct experience, and 2)  {\bf providing complete and usable third-party solutions in this space is non-trivial}---or the cloud vendors who already had internal versions of this would have already done so. We speculate that this relates to the extra challenges introduced by EGML, and believe this area is primed for disruptive research, as we discuss in more details in \S~\ref{sec:problems}.

\vspace{2mm}
{\noindent \em \large Research in the ML community.} Research in the area of machine learning systems is plentiful, and too large an area to do justice to here. Instead, we focus on major trends that we observed over the last 15~years that influenced our thinking on the intersection of machine learning and data management systems.

After an initial focus on algorithms, the ML community has concentrated (in rough order of appearance) on: 1) systems for training, 2) systems for scoring, 3) AutoML solutions, and 4) responsible AI. The systems for training area was initially dominated by big-data extensions \cite{mahout,sparkmllib} first and HPC-based solutions \cite{horovod}, and later by parameter servers with bounded staleness \cite{boundedstaleness}. More recent attempts such as ML.NET~\cite{ml.net} and TFX~\cite{tfx} have borrowed more profoundly from the dataflow/database literature to build ML-native solutions. Systems for scoring were vastly ignored until systems such as Clipper~\cite{clipper}, Pretzel~\cite{pretzel} and TensorFlow serving~\cite{tfserving} came to fruition. They draw heavily from streaming systems and optimizing compilers in their design.  As ML adoption began to broaden, AutoML solutions began to appear~\cite{automl_book}.  Lately, interest in bias, fairness and responsible use of machine learning is exploding, though only limited solutions exist. This aligns with feedback from enterprise customers (i.e., ``automate it, and don't get me~sued'').

We conclude that {\bf data platforms play a key role to achieve fast and reliable training and scoring, and that explicit metadata management and provenance tracking are foundational for responsible AI and AutoML solutions}.

\section{Open Problems \& Advances}\label{sec:problems}

The vision for EGML we presented is an exciting one and presents many challenging problems. We summarize some key challenges below and present some of our ongoing work.
We focus on two categories that require attention from the DB community and are not well understood: 1) the systems support required to go from a trained model to decisions, and 2) data management for ML.

\subsection{From Model to Decision: Inference} 
\label{sec:inference}

Much attention has been given to learning algorithms and efficient model training, but models only have value insofar as they are used for inference, to create insights and make decisions. This typically involves a complicated setup of containers for deploying the trained model (as executable code), with applications invoking them via HTTP/REST calls.  Further, the containerized code often extends model inference with the implementation of complex application-level policies.

While this containerized approach offers a desirable decomposition of the problem between models and the applications using them, it has significant drawbacks: (1) Many applications use more than one model, with each model applied to the outcome of some (potentially different) data processing step. These assemblies of models and preprocessing steps should be updated atomically. (2) It seems unlikely that this solution will fit the scenarios emerging from the millions of applications we expect in this space (e.g., latency-sensitive decisions and large batch predictions are poorly served). (3) Mixing application-level policies and inference logic makes it hard to separate and measure the impact of the two.

We believe that models should be represented as first-class data types in a DBMS. This will naturally address (1) by allowing database transactions to be used for updating multiple deployed models.  To address (2), we believe inference/scoring should be viewed as an extension of relational query processing, and argue for moving model inference close to the data and performing it in-DBMS, without external calls for common types of models. Naturally, this calls for a separation of inference from application-level logic. We present a clean framework for (3) after we briefly summarize our early results on in-DBMS inference. A more in-depth discussion on inference appears in~\cite{raven}.

\vspace{2mm}
{\noindent \em \large In-DBMS inference.} While in-DBMS inference appears desirable, a key question arises: {\em Can in-DBMS model inference perform as well as standalone dedicated solutions?}

To this end, several recent works~\cite{madlib,lara,laradb,levelheaded} in the database community explore how linear and relational algebra can be co-optimized. To carry this investigation further, we integrated ONNX Runtime~\cite{onnx-runtime} within SQL Server and developed an in-database cross-optimizer between SQL and ML to enable optimizations across hybrid relational and ML expressions~\cite{raven}. Further, we observe that practical end-to-end prediction pipelines are composed of a larger variety of operators (e.g., featurizers such as text encoding and models such as decision trees) often assembled in Python. We leverage static analysis to derive an intermediate representation (IR) amenable to optimization. The list of optimizations we have been exploring is therefore more comprehensive than prior work, and includes the following:

\begin{itemize}[noitemsep,topsep=0pt,parsep=0pt,partopsep=0pt]
\item predicate-based model pruning, which uses selections from the SQL subsection of the query to simplify the ML subsection; 
\item model-projection pushdown, which automatically prunes (projects out) unused input feature-columns exploiting model-sparsity; 
\item model clustering that compiles simplified models for different parts of the input based on data statistics; 
\item model inlining, which transforms ML operators to relational ones (similar in spirit to~\cite{froid});
\item physical operator selection based on statistics, available runtime (SQL/ONNX/Python UDFs~\cite{spexternalsql}) and hardware (CPU,~GPU).
\end{itemize}
\vspace{2mm}

\begin{figure}[t!]
    \includegraphics[width=0.32\textwidth]{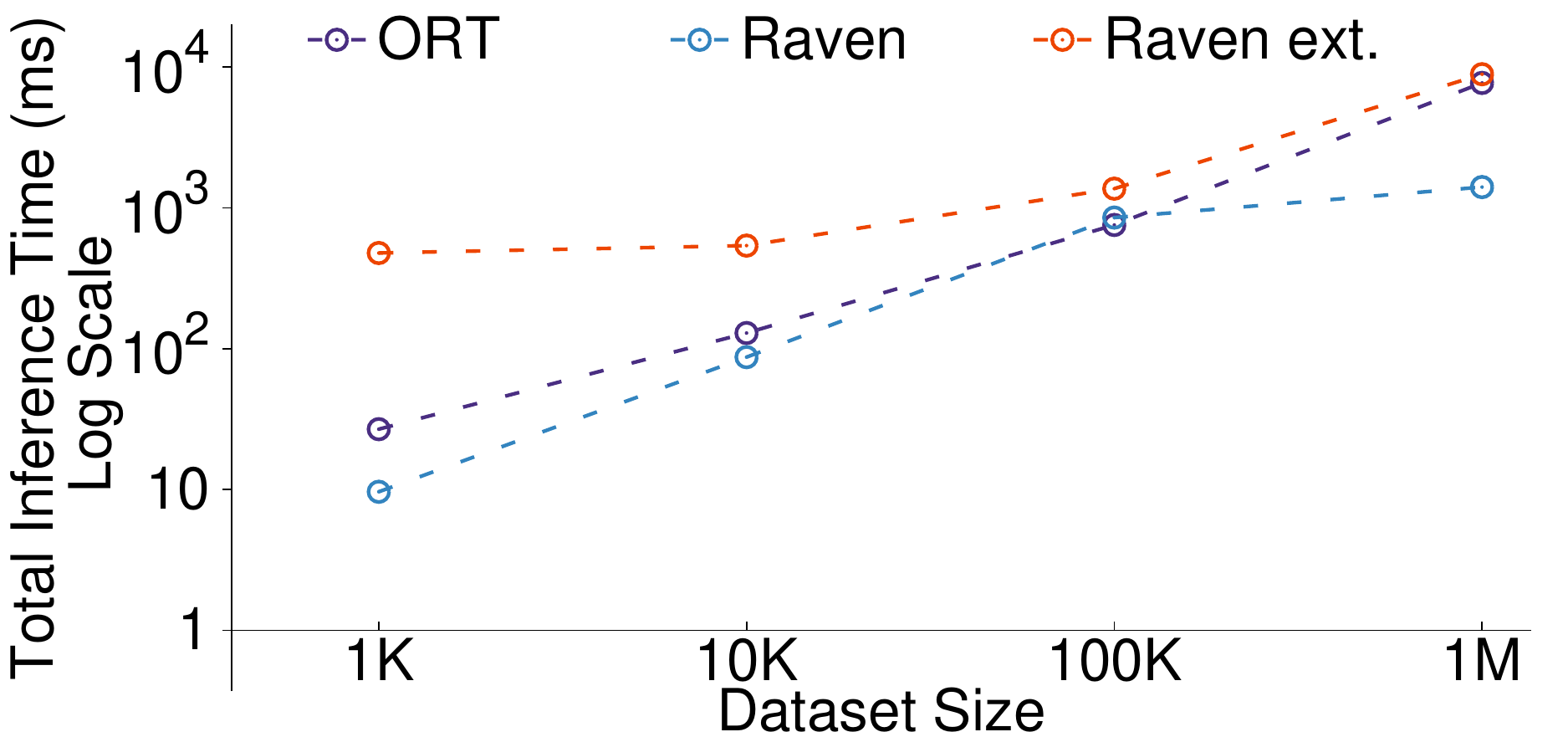}
	\hspace{0ex}\includegraphics[width=0.15\textwidth,trim={2.6cm 0 0 0},clip]{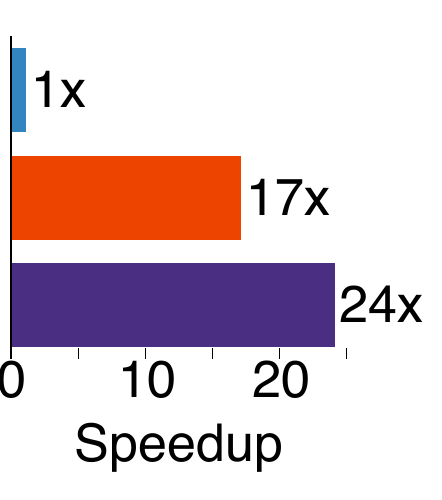}
        \vspacebfigure
	\caption{Benefits of integrating ONNX Runtime in SQL Server (left) and of cross-optimizations (right---red bar shows speedup due to model inlining, blue bar also includes predicate-based model pruning).
	\label{fig:inference-eval}}
        \vspaceafigure
\end{figure}

In Figure~\ref{fig:inference-eval} we present two key results: (1) performance of integrating ONNX Runtime (ORT) within SQL Server, both through in-process execution (Raven) and out-of-process (Raven ext.); and (2) benefit of cross-optimizations, including model inlining and predicate-based model pruning. The results show that SQL Server integration provides up to $5.5\times$ over standalone ORT (due to automatic parallelization of the inference task in SQL Server) and up to $24\times$ from our combined optimizations.  Early results indicate that in-DBMS inference is very  promising.

\vspace{2mm}
{\noindent \em \large Bridging the model-application divide.}
ML applications need to transform the model predictions into actionable decisions in the application domain. However, the mathematical output of the model is rarely the only parameter considered before a decision is made. In real deployment scenarios, business rules and constraints are important factors that need to be taken into account before any action is taken. As a concrete example, we have built models to automate the selection of parallelism for large big data jobs to avoid resource wastage (in the context of Cosmos clusters~\cite{morpheus}). While models are generally accurate, they occasionally predict resource requirements in excess of the amounts allowed by user-specified limits. Business rules expressed as policies then override the model.

Obviously, business rules and requirements can vary between different applications and environments. To that end, we employ a generic and extensible module~\cite{floratou2017dhalion} that takes as input user-defined \textit{policies} which introduce various business constraints on top of EGML workloads. The module continuously monitors the output of the ML models and applies the specified policies before taking any further action in the application domain. It also maintains the system state and actions taken over time allowing us to easily debug and explain the system's actions. Finally, it makes sure that the actions happen in a transactional way, rolling back in case of failures when needed.  Overall this closes the loop between model and application, providing us visibility necessary for both debugging and end-to-end accountability.

Next, we discuss the requirements for managing data for ML.

\subsection{Data Management for ML} 
\label{sec:datamanagement}

\vspace{2mm}
{\noindent \em \large Data Discovery, Access and Versioning.} One of the main challenges ML practitioners face today revolves around data access and discovery. Training data commonly contains tabular data, but also images, video or other sensor data. This gives rise to a predominantly file-based workflow. Only a small fraction of the $>4$~million notebooks we analyzed makes use of a database access library. This is surprising, as the vast majority of the pipelines ultimately use Pandas~\cite{pandas}, a structured DataFrame, to interact with this data. This state-of-the-art is deeply unsatisfying: {\em Data Discovery support is virtually non-existent }. This is especially troubling as data augmentation is one of the best strategies to improve a model.

Worse, \emph{data versioning is largely unsolved in this paradigm:} A model is the result of both its training code and the training data. Both need to be versioned. And file versioning technologies fail to address key needs of data versions; they often can only represent a deletion via a history rewrite. More fundamentally, files are not the atomic unit of training data: an individual data point may be stored in a file, but equally likely, many files represent one data point, or one file contains many data points.

Hence, we believe that  {\bf there is an open need for queryable data abstractions, lineage-tracking and storage technology that can cover heterogenous, versioned, and durable data.}

\vspace{2mm}
{\noindent \em \large Model Management.}We have argued that ML models are software artifacts created from data, and must be secured, tracked and managed on par with other high-value data. DBMSs provide a convenient starting point thanks to their support for enterprise-grade features such as security, auditability, versioning, high availability, etc. To be clear, we are not suggesting that all data management needs to be inside a relational DBMS; indeed, we see a trend towards comprehensive data management suites that span all of a user's data across one or more repositories.  Our point is that managing models should be treated on par with how high-value data is managed, whether in a DBMS (the most widely available option currently) or in emerging cross-repository managed environments.  

 \begin{figure}
	\centering\includegraphics[width=\columnwidth]{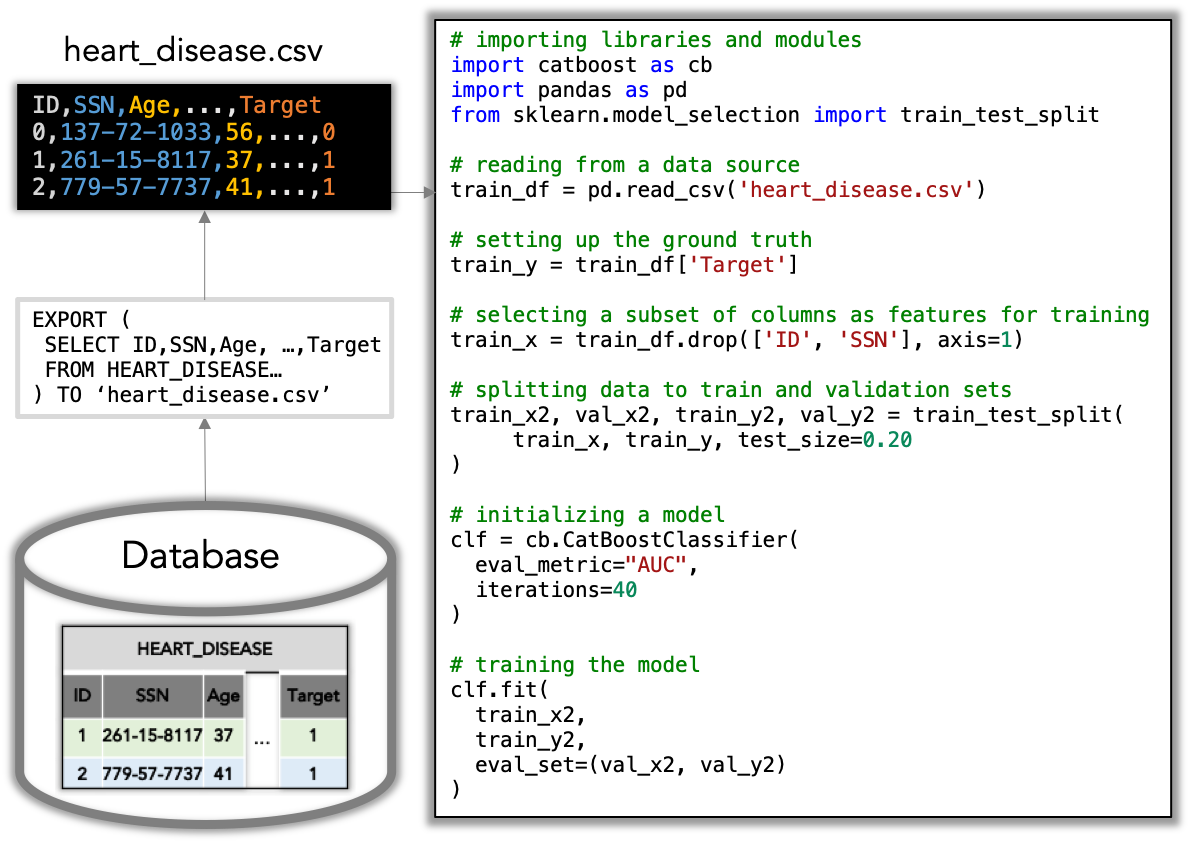}
        \vspacebfigure
	\caption{Data science pipeline}
	\label{fig:ds}
        \vspace{-1em}
\end{figure}

\vspace{2mm}
{\noindent \em \large  Model Tracking and Provenance.} Models are software of consequence. Their genesis needs to be tracked. To achieve that, the full \emph{provenance} of a model must be known for debugging/auditing.

We need to capture not only the code that trained the model, but also the (training) data that went into it, together with its full, tamper-proof lineage. There are multiple industry efforts to capture the inner training loop of this lineage~\cite{mlflow, kubeflow}. This must be expanded to the full lineage, and also automated to achieve the scale we expect.  

Figure~\ref{fig:ds} shows an example of a data science pipeline. The data scientist first selects a subset of patient-related data stored in a relational database and saves them in a CSV file. This file can then be accessed by a Python script to train a machine learning model. This pipeline involves different tools and languages (SQL, Python) and data formats (SQL tables, flat files).

 Tracking  provenance across such pipelines can enable various important applications such as model linting, compliance testing, and impact analysis. In this particular example, the label used to train the model (TARGET column) is also part of the feature set. Automatically analyzing the Python script to identify the provenance relationships between the columns of the dataset and the trained model can help identify such mistakes. Similar techniques can also be used to detect compliance violations such as usage of PII information by a machine learning model. In our example, the dataset contains sensitive patient data that include the SSN and ID columns. However, these columns are dropped from the Python script before training the machine learning model, and as a result the pipeline is compliant with the regulations. Finally, automatic tracking of provenance can help identify the set of applications that are affected by changes in the original data sources. For example, one can use such information to detect which machine learning models will be impacted when a column of the dataset in the database is dropped because it contains corrupted data. 

Enabling such applications by automatically capturing provenance information across such pipelines is challenging:

\begin{itemize}

\item \sstitle{C1. Provenance data model.} Data elements in EGML workloads are polymorphic (e.g., tables, columns, rows, ML models, and hyper-parameters) with inherent temporal dimensions (e.g., a model may have multiple versions, one for each re-run of a training pipeline). As such, and in contrast to traditional data models of provenance over DBMSs, EGML workloads dictate polymorphic and temporal provenance data models. Such data models are hard to design, capture, maintain, and query. 

\item \sstitle{C2. Provenance capture.} EGML workloads typically span multiple systems and runtimes (e.g., a Python script may fetch data from multiple databases to train a model). These systems might have different architecture and programming constructs (e.g., declarative vs. imperative interfaces). Extracting a meaningful provenance data model in this setting requires different capture techniques tailored specifically to each system and runtime.

\item \sstitle{C3. Provenance across disparate systems.} Even if we capture provenance on top of each system and runtime in isolation, we still require to combine this information across systems (e.g., if we change a column in a database, models trained in Python that depend on this column may need to be invalidated and retrained). Hence, EGML workloads require protocols for consolidating and communicating the provenance information across systems.

\end{itemize}

\sstitle{Our initial solution.} Our solution consists of three major modules: the \textit{SQL Provenance module}, the \textit{Python Provenance module} and the \textit{Catalog}. The Catalog (we use Apache Atlas~\cite{atlas}) stores all the provenance information and acts as the bridge between the SQL and the Python Provenance modules. It allows us to capture end-to-end provenance across different systems---hence, provides a principal way to address \textbf{C3}. Next, we present the high-level architecture and some preliminary experimental results (full papers with the detailed description of the system design and associated algorithms behind the SQL and Python Provenance modules are under preparation).

\vspace{2mm}
\sstitle{Provenance in SQL.} Our SQL provenance module currently focuses on capturing coarse-grained provenance under two modes, traditionally referred to as eager and lazy. Under eager provenance capture, given a query, the module parses it to extract coarse-grained provenance information (i.e., input tables and columns that affected the output, with connections modelled as a graph). Under lazy provenance capture, the module gets as input the query log of the database and constructs the provenance data model, only this time by accounting the whole query history. Under both modes, the module populates the Catalog accordingly. 

To scale across databases, the parsing module utilizes Apache Calcite~\cite{calcite} that provides parsers and adapters across databases---hence, provides us a way towards addressing \textbf{C2}. For cases where Apache Calcite cannot parse queries, we specialize to the parser of the corresponding engine. Furthermore, note that all data stored in the Catalog is versioned (e.g., an \texttt{INSERT} to a table results in a new version of the table in the provenance data model)---hence, we address the temporal aspect of \textbf{C1}.  The table below shows the provenance capture performance (latency and provenance graph size) for queries generated out of all templates in TPC-H and TPC-C:
\vspace{-3mm} %
\begin{table}[h]
\centering
\small
\begin{tabular}{| c|c|c|c| } 
 \hline
 \textbf{Dataset} & \textbf{\#Queries} & \textbf{Latency} & \textbf{Size(nodes+edges)}\\
 \hline
TPC-H & 2,208 & 110s & 22,330\\ \hline
TPC-C & 2,200 & 124s & 34,785 \\ 
 \hline
\end{tabular}
\label{t:sqlperf}
\end{table}
\vspace{-1.5mm}

These early findings indicate that a) the per-query capture latency can be significant and b) the provenance data model can become substantially large in size (e.g., a table with as many versions as the number of insertions that have happened to it). For these reasons, we develop optimized capture techniques, through compression and summarization, which are essential towards addressing \textbf{C1}.

\vspace{2mm}
\sstitle{Provenance in Python.} The Python provenance module parses scripts and automatically identifies the lines of code that correspond to feature extraction and model training using a combination of standard static analysis techniques and a knowledge base of ML APIs that we maintain. Through this process, we are able to identify which Python variables correspond to models, hyperparameters, model features and metrics. We can also track the transformations performed on these variables and eventually connect them with the datasets used to generate training data. The Python provenance module accesses the Catalog to collect the output of the SQL provenance module and eventually connect the datasets used in the Python scripts to the columns of one or more DBMS tables. 
\vspace{-1.5mm}
\begin{table}[h]
\centering
\small
\begin{tabular}{| c|c|c|c| } 
 \hline
 \textbf{Dataset} & \textbf{\#Scripts} & \textbf{\%Models} & \textbf{\%Training Datasets}\\ 
  &  & \textbf{ Covered} & \textbf{Covered}\\ \hline
Kaggle & 49 &$95\%$ & $61\%$ \\ \hline
Microsoft & 37 & $100\%$ & $100\%$ \\ 
 \hline
\end{tabular}
\label{tab:pythoncoverage}
\end{table}
\vspace{-1.5mm}

The above table table shows the coverage currently achieved by the provenance module on the Kaggle dataset~\cite{kaggle} and a Microsoft internal dataset of scripts deployed in production. In this experiment, we evaluate how often the module identifies correctly ML models and training datasets in the Python scripts.

\section{Conclusion and Call to Action}
\label{sec:conclusions}

We live in interesting times. Database architectures are undergoing major transformations to leverage the elasticity of clouds, and a combination of increased regulatory pressures and data sprawl is forcing us to rethink data governance more broadly.  Against this backdrop, the rapid adoption of ML in enterprises raises foundational questions at the intersection of model training, inference and governance, and we believe the DB community needs to play a significant role in shaping the future.

\section*{Acknowledgements}
We thank Doris Xin, Mohammad Hossein Namaki and Yi Zhang for their contributions in Sections~\ref{sec:inference} and \ref{sec:datamanagement}---full-length papers on each system contribution are ongoing.

\bibliographystyle{abbrv}
\bibliography{references}

\end{document}